\documentclass[11pt,twoside]{article}

\usepackage{asp2006}
\usepackage{epsf}
\usepackage{psfig}
\usepackage{lscape}
\usepackage{graphicx}

\begin{document}
\title{Hot GRB-selected Submillimeter Galaxies}   
\author{M.~J.~Micha{\l}owski, J.~Hjorth,  J.~M.~Castro Cer\'{o}n and D.~Watson} 
\affil{{Dark Cosmology Centre, Niels Bohr Institute, University of Copenhagen}, Juliane Maries vej 30, 2100 Copenhagen \O, Denmark}    

\begin{abstract} 
Using  detailed spectral energy distribution fits we present evidence that 
submillimeter- and radio-bright gamma-ray burst host galaxies are hotter counterparts to submillimeter galaxies. This hypothesis makes them of special interest since hotter submm galaxies are difficult to find and are believed to contribute significantly to the star formation history of the Universe.
\end{abstract}


\section{Introduction}

It is widely accepted that cold and luminous submillimeter (submm) galaxies (SMGs)  are dominant contributors to the star formation history of the Universe at redshifts $z\sim2-3$ \citep{chapman05}. 
On the other hand, the host galaxies of long-duration gamma-ray bursts \citep[GRBs, originating in the collapses of very massive stars, e.g.][]{hjorthnature}   are found to be subluminous  \citep{fruchter06} and low-mass \citep[][]{castroceron06}.  
Four of them (GRBs 980703, 000210, 000418 and 010222) have been firmly detected in submm and/or radio (\citeauthor{tanvir} \citeyear{tanvir}; \citeauthor*{bergerkulkarni} \citeyear{bergerkulkarni}; \citeauthor{berger} \citeyear{berger}). 
In this paper we  discuss the possibility that these submm-bright GRB hosts may represent the hotter counterparts of SMGs. 
For details see \citet{michalowski07} and \citet{michalowski06}.

\section{SED Modeling and Results}

In order to model the spectral energy distributions (SEDs) of GRB hosts we used the GRASIL
software  \citep{silva98}.
It is a numerical code that calculates the spectrum of
a galaxy by means of a radiative transfer method, applied to photons produced by a stellar population,
and reprocessed by dust. 

\begin{figure}
\begin{center}
\includegraphics[width = .55\textwidth]{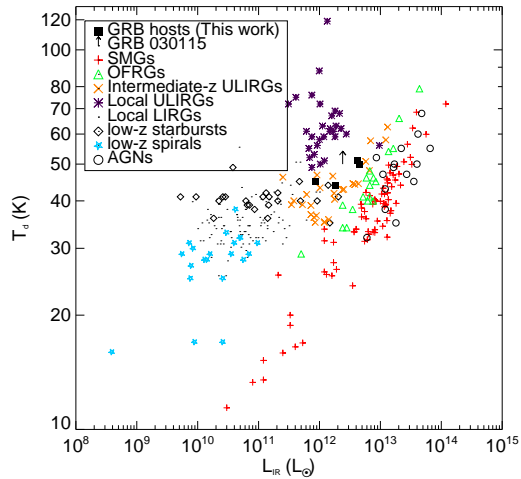}
\end{center}
\caption{Dust temperature as a function of infrared  luminosity. GRB hosts discussed here 
and GRB 030115 
are compared with different  galaxy samples  \citep[see][and references therein]{michalowski07}. 
Note that for GRB 030115 and OFRGs a lower limits to dust temperature are shown.
GRB hosts seem to reflect the properties of intermediate-$z$ ULIRGs, the bright end of starburst galaxies and OFRGs --- the candidates for hotter counterparts of SMGs.}
\label{fig:T_L}
\end{figure}

In Figure~\ref{fig:T_L} we compare the total infrared luminosities  and dust temperatures of GRB hosts (derived from the SED fits) with well-studied galaxies both local and at high-$z$. 
It is apparent that GRB hosts are hotter than SMGs with the same luminosity. This gives a hint that GRB events may pinpoint a population of ultraluminous infrared galaxies (ULIRGs) at high redshifts with dust hotter than in typical SMGs. 
The search for such galaxies is important because they likely contribute to the star formation history at the same level as SMGs.

We note that the majority of the galaxies shown in Figure~\ref{fig:T_L}  also have higher dust temperatures compared to SMGs. However, all of them are local galaxies, so cannot be considered as counterparts of high-redshift SMGs and their submm emission has been detected only because of their proximity.

GRB hosts may be consistent with a population of optically faint radio galaxies (OFRGs) having similar infrared luminosities and (likely) temperatures. 
Although the majority of OFRGs lie at $z\sim2$ \citep{chapman04}, some of them are within the redshift range of the GRB hosts discussed here (\mbox{$z=0.85-1.48$}).
OFRGs have been suggested to be hotter counterparts of SMGs \citep{chapman04}, so the same may be true for GRB hosts.

\acknowledgements 
The Dark Cosmology Centre is funded by the Danish National Research Foundation.


\end{document}